\documentclass[12pt,preprint]{aastex}

\shorttitle{Magnetic Flux Emergence in the Quiet Sun}
\shortauthors{Centeno et al.}

\begin{document}

\title{Emergence of Small-Scale Magnetic Loops in the Quiet Sun Internetwork}

\author{R. Centeno, H. Socas-Navarro, B. Lites, M. Kubo}
\affil{High Altitude Observatory (NCAR)\footnote{The National Center for Atmospheric Research is sponsored by the National Science Foundation}, Boulder CO 80301, USA}
\author{Z. Frank, R. Shine, T. Tarbell,  A. Title}
\affil{Lockheed Martin Space and Astrophysics Laboratory, Palo Alto, CA 94304, USA}
\author{K. Ichimoto, S. Tsuneta, Y. Katsukawa, Y. Suematsu }
\affil{National Astronomical Observatory of Japan, Tokyo, Japan}
\author{T. Shimizu}
\affil{Japan Aerospace Exploration Agency, Tokyo, Japan}
\and
\author{S. Nagata}
\affil{Kwasan and Hida Observatories, Kyoto University, Japan}

\email{rce@hao.ucar.edu}

\begin{abstract}
We study the emergence of magnetic flux at very small spatial scales
(less than 2 arcsec) in the quiet Sun internetwork. To this aim, a
time series of spectropolarimetric maps was taken at disk center using
the instrument SP/SOT on board Hinode. The LTE inversion of the full Stokes
vector measured in the Fe {\sc i} 6301 and 6302 \AA\ lines allows us to
retrieve the magnetic flux and topology in the region of study.
In the example presented here, the magnetic flux emerges within a granular
structure. The horizontal magnetic field appears prior to
any significant amount of vertical field. As time goes on, the traces of the
horizontal field disappear while the the vertical dipoles drift -carried by
the plasma motions- towards the surrounding intergranular lanes.
These events take place within typical granulation timescales.

\end{abstract}


\section{Introduction}

The magnetic flux emerges on many different spatial scales on the
solar surface in the form of bipolar regions connected by magnetic loops.
Time scales span from weeks (or even months) for tens-of-Mm active regions,
to several minutes for very small-scale magnetic structures at the resolution 
limit of modern solar telescopes. 

\noindent The nature of internetwork (IN) magnetic fields is now being debated 
heatedly in Solar Physics. While some authors defend the
idea of strong kG field strengths associated with small filling factors
(see e.g. \cite{jorge00}, \cite{itahiza}), other work suggests the 
predominance of
weak magnetic fields (\cite{lin95}, \cite{linrimmele99},
\cite{khomenko03}) in the range of 300-500 G.
\cite{manso04} have reported the detection of an ubiquotous 
unresolved turbulent magnetic field by measuring the Hanle effect in the 
scattering polarization signal of the Ti {\sc i} 4535 \AA\ line.
Regardless of their nature, the emergence of these small-scale IN 
magnetic fields is most likely related to the convection process underneath
the photosphere (\cite{depontieu02}, \cite{cheung07}). The horizontal internetwork fields
(HIFs) reported by \cite{lites96}, with typical sizes of 1" and lifetimes
of $\sim 5$ minutes, suggest that small magnetic loops are being advected
towards the surface by the upward motion of the plasma inside the granules.
If HIFs are indeed emerging flux, \cite{lites96} estimated that the rate 
of magnetic flux driven to the surface by this mechanism is greater than 
the rate of flux emergence in bipolar sunspot regions averaged over the
whole solar cycle.
 
The measurement of the full topology of a magnetic loop requires
accurate spectropolarimetric 2-D maps of the four Stokes parameters, 
with high S/N ratio, high spatial resolution and consistent seeing
conditions (\cite{marian07}). The Spectro-Polarimeter (SP)
(\cite{lites01}) of the Solar 
Optical Telescope on board Hinode \cite{kosugi07} meets all these 
requirements, and the 
space-based observations guarantee the total absence of seeing-induced 
cross-talk among the Stokes measurements.

Here we present a clear evidence of the emergence of a small-scale IN magnetic
loop in the quiet Sun photosphere obtained from measurements done with
SP on board Hinode. We followed the event in time from 
the moment it showed a measurable magnetic signal until it was fully emerged 
and developed.

\section{Observations}

The data presented here were obtained with Hinode's SP on March 10, 2007. 
They are a part of a 5-hour-long time series of spectropolarimetric maps 
(4" wide and 82" long) taken on the quiet Sun disk center with a time
cadence of approximately 2 minutes per map.
Each map was constructed from 25 consecutive positions of the spectrograph
slit, with an integration time of 4.8 s per position. The slit was set to 
scan westwards at 0.16" per step, ending in a series of 4" wide maps 
with a spatial resolution of $\sim$0".32. 
The spectral region measured by SP contains two photospheric Fe {\sc i} 
lines (at $\lambda$6301.5 and $\lambda$6302.5 \AA). The full Stokes 
($I$, $Q$, $U$ and $V$) profiles were obtained for every position along the 
slit with a spectral sampling of 21.5 m\AA\ per pixel. The noise level
in the continuum polarization of $\sim 1.2\cdot10^{-3}I_c$,
together with the high spatial resolution and the absence of seeing induced
crosstalk makes it possible to detect the small-scale weak magnetic signals
we are looking for.

\begin{figure}[t]
\includegraphics[scale=.40]{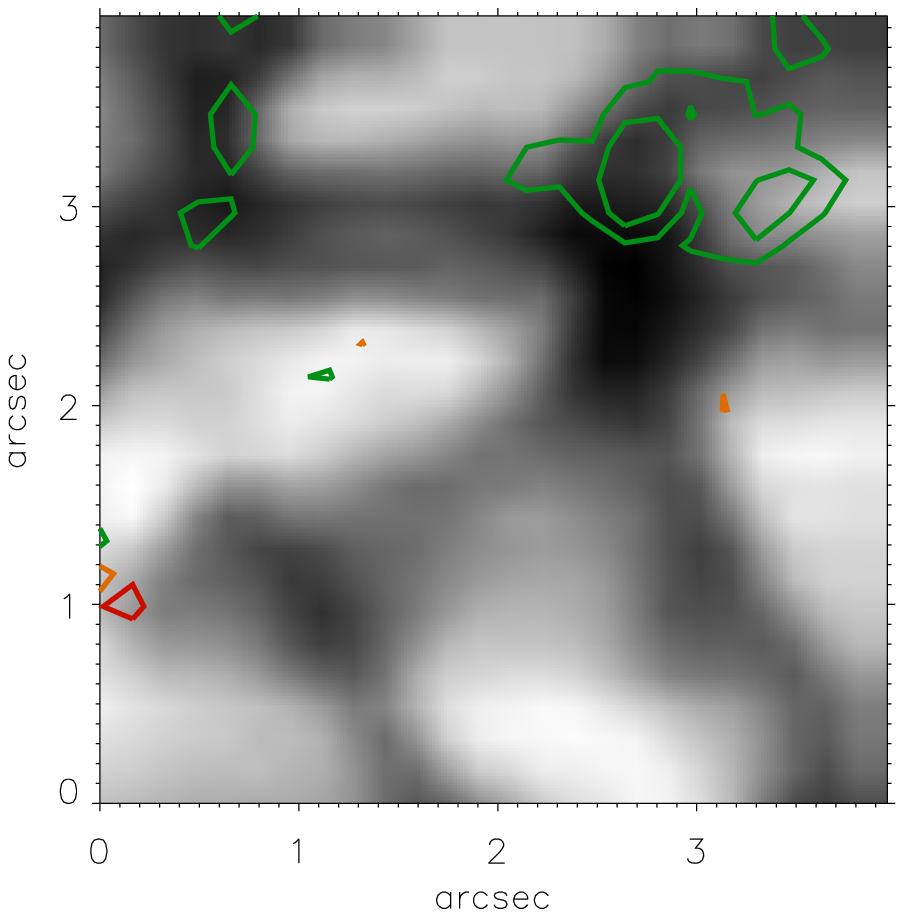}
\includegraphics[scale=.40]{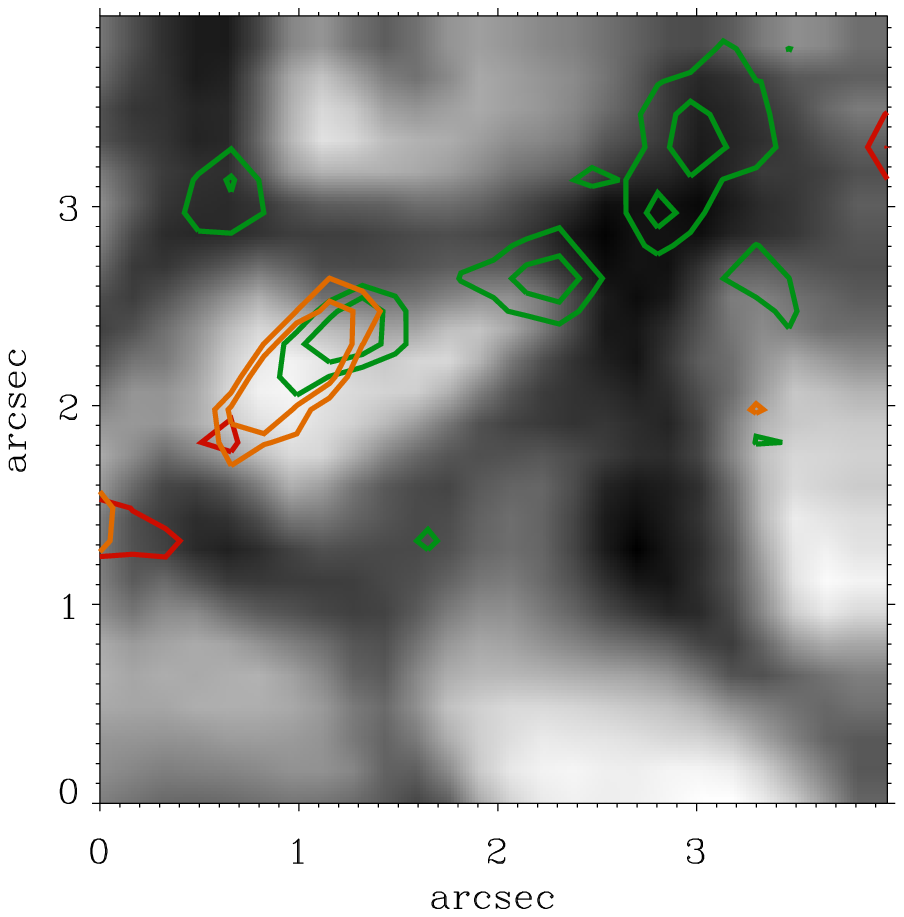}
\includegraphics[scale=.40]{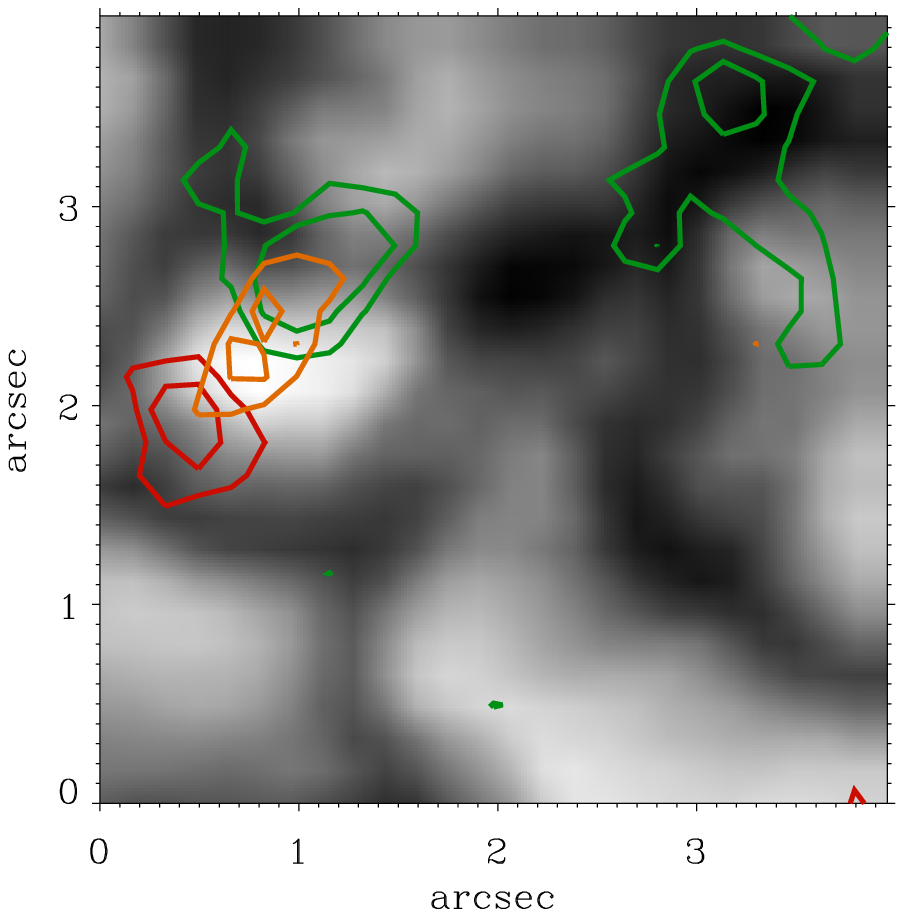}
\includegraphics[scale=.40]{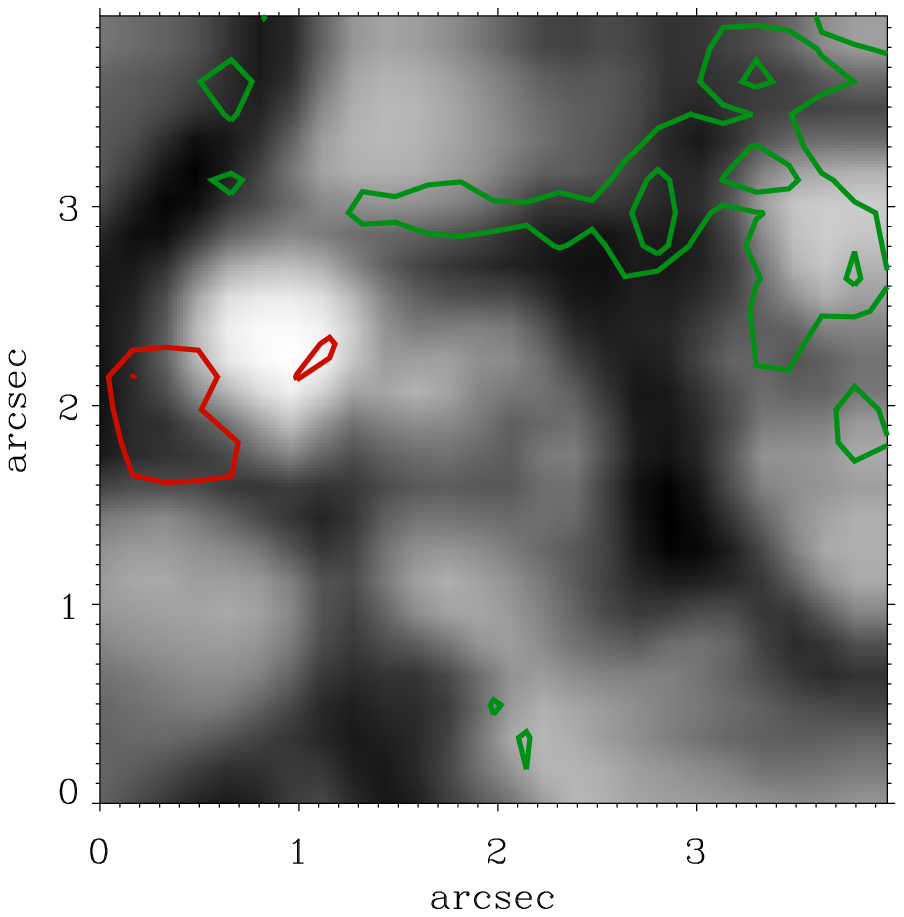}
\caption{4"x4" box shown in 4 consecutive snapshots of the time series, 
separated by 125 s. W/B background shows the granulation pattern measured 
by the integrated continuum intensity. Red, green and orange contours 
represent positive circular, negative circular and linear net polarization 
signals, respectively. The outer contour levels correspond to $3\sigma$ in
the case of Stokes $V$ and 6$\sigma$ for the net linear polarization. 
Note the clear spatial correlation between the emerging magnetic flux and
the granule location.}
\end{figure}

In order to find flux emergence events in the quiet Sun photosphere we
searched the time sequence of magnetograms (constructed from the information
contained in the Stokes $Q$, $U$ and $V$ profiles) for isolated regions of 
emerging flux within the IN, avoiding the persistent strong magnetic flux 
concentrations of the network areas.

\noindent Fig. 1 shows an example of one of these events happening throughout 
4 consecutive snapshots (separated in time by 125 s) of the data-set in 
a 4"x4" box. 
The gray-scale background shows the integrated continuum intensity, which 
reveals the underlying photospheric granulation pattern. The overplotted 
contours show the areas with non-negligible polarization signals. While the
orange contours show the net linear polarization signals ($\sqrt{Q^2+U^2}$),
the green and red ones represent the negative and positive circular 
polarization, respectively.
In the first snapshot, the locations of all significant magnetic signals are 
contained within the intergranular regions, in agreement with the 
widely accepted picture of 
the magnetic flux tubes being advected to the intergranular lanes. 
In the second snapshot, a non-negligible amount
of linear polarization (horizontal magnetic field), flanked by two weak 
opposite polarity circular signals (vertical magnetic field), emerges 
from inside a granule in the center-left region of the box. This configuration
is consistent with the topology of a small magnetic loop with its
footpoints standing at each side of the region of linear polarization 
signal. 
Two minutes later, this structure becomes more developed, spanning over a
direction that forms an angle of $\sim60$ degrees with respect to the 
horizontal axis (E-W direction) of the box.
In the last snapshot we can see how the linear signal has disappeared while 
the remaining longitudinal magnetic flux has been swept towards 
the surrounding intergranular lanes, presumably driven by the horizontal 
plasma motions.

\noindent Note that the magnetic flux contained within the intergranules in 
the first snapshot is trapped there throughout the whole sequence. It is 
only the new emerging flux that appears co-spatial with granule locations.

\section{Magnetic flux density and field topology}

\begin{figure}[!t]

\includegraphics[scale=.38]{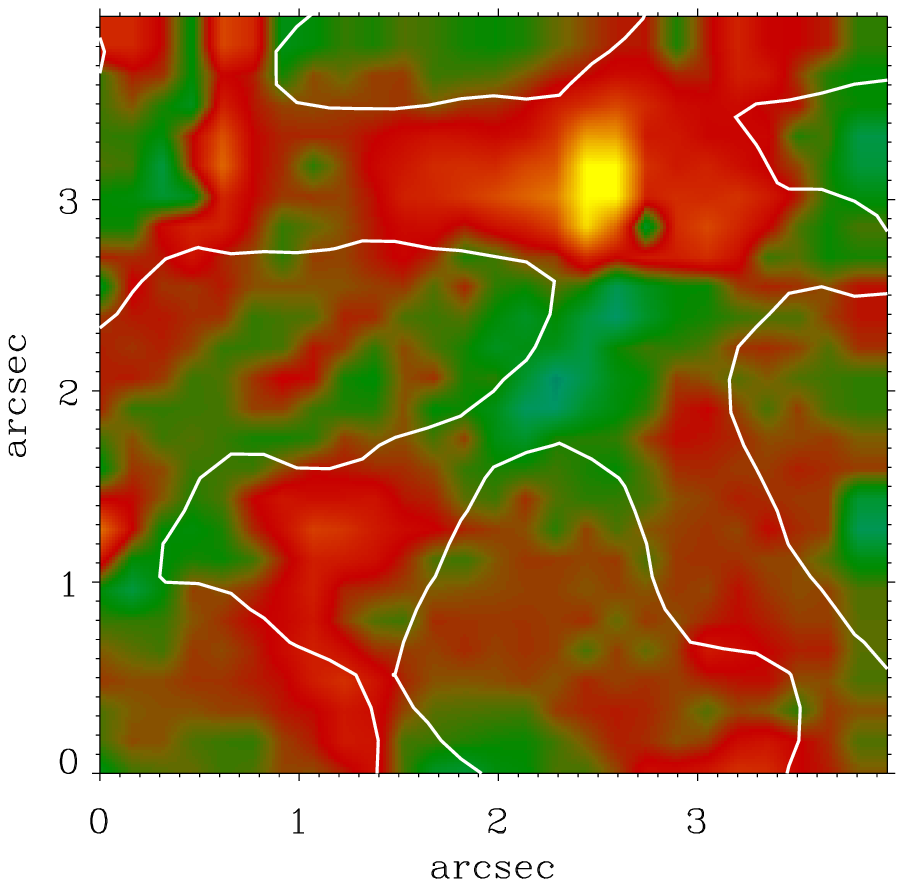}
\includegraphics[scale=.38]{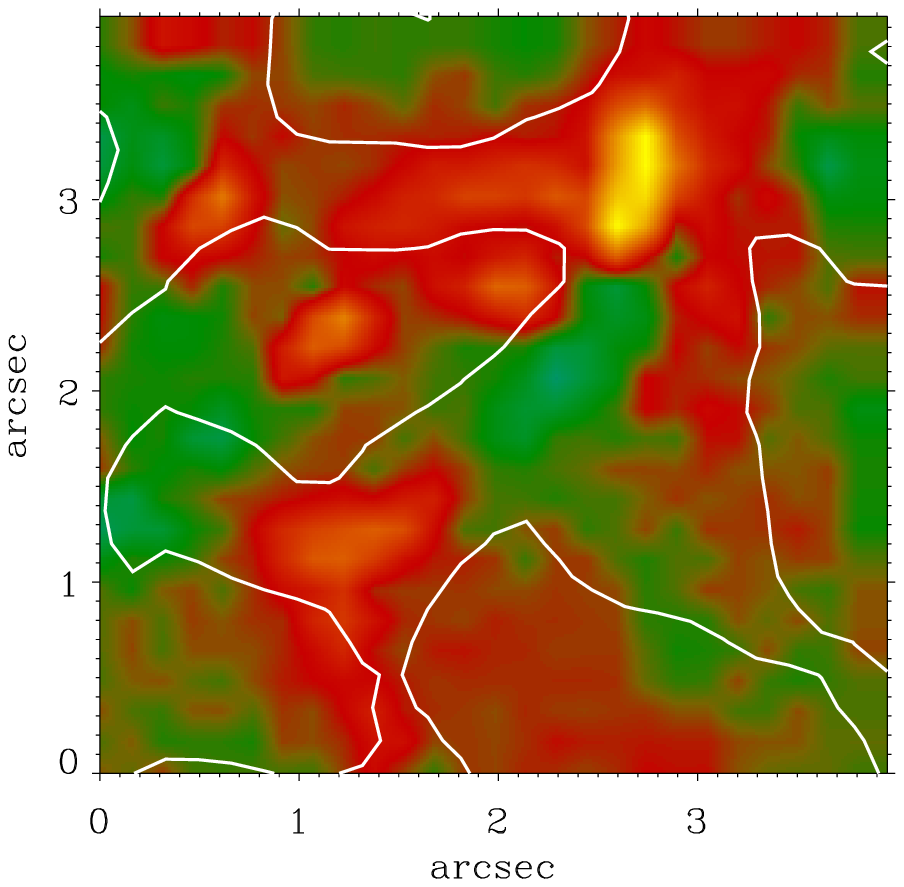}
\includegraphics[scale=.38]{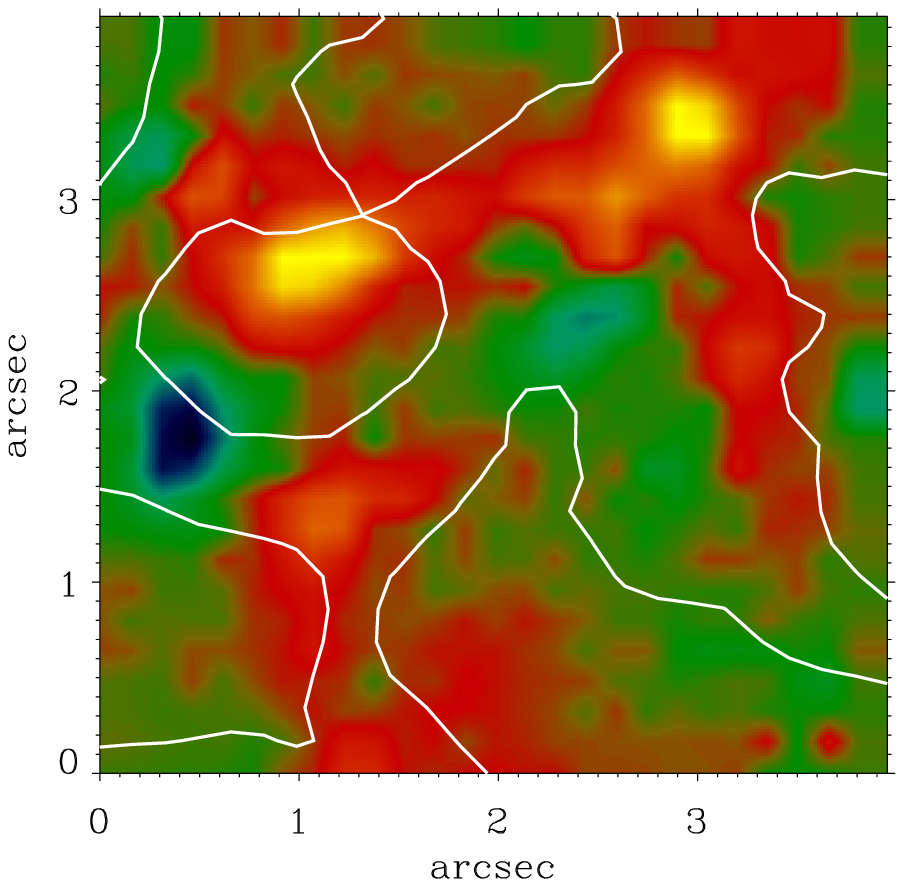}
\includegraphics[scale=.38]{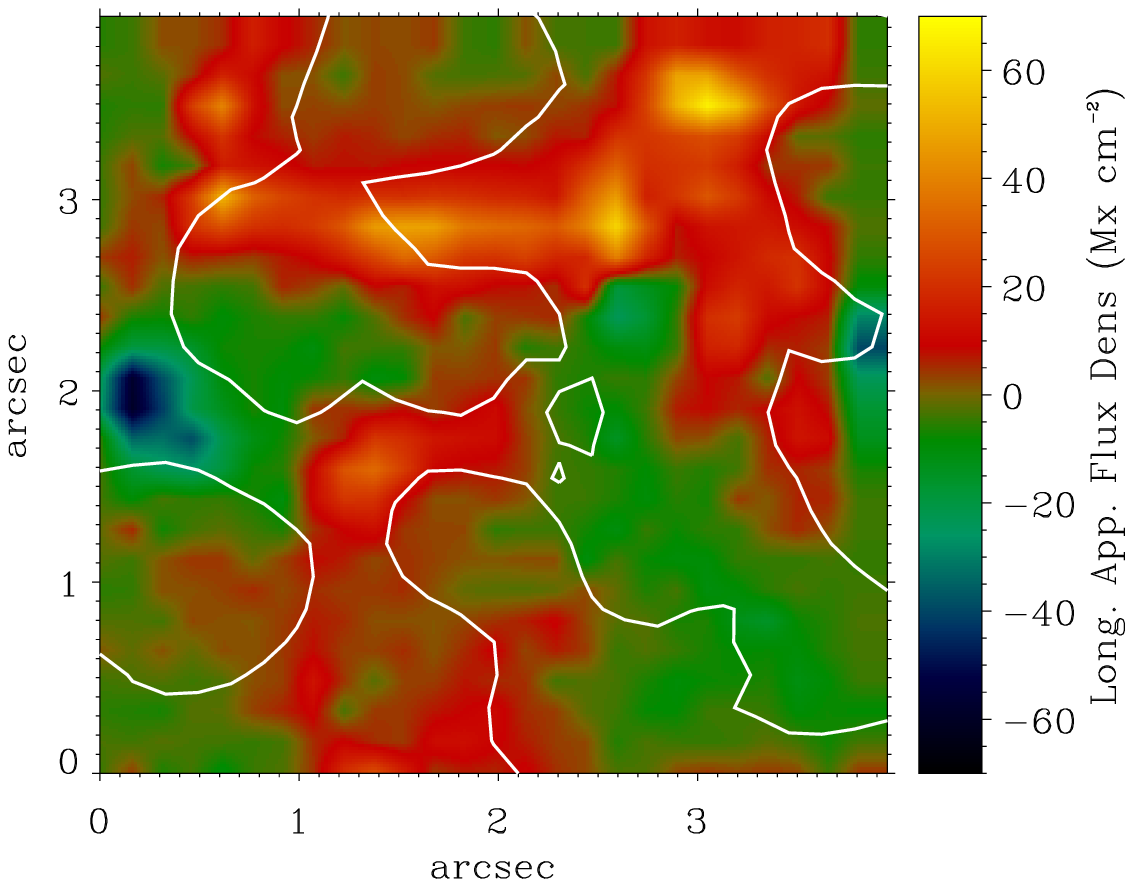}\\
\includegraphics[scale=.38]{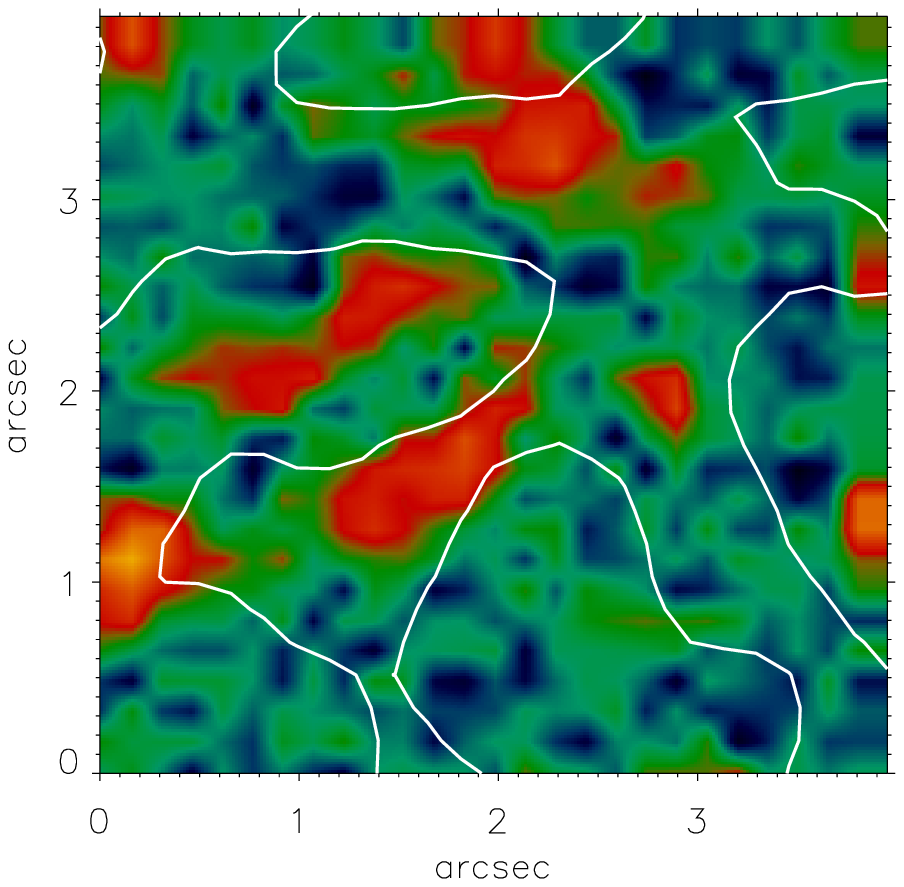}
\includegraphics[scale=.38]{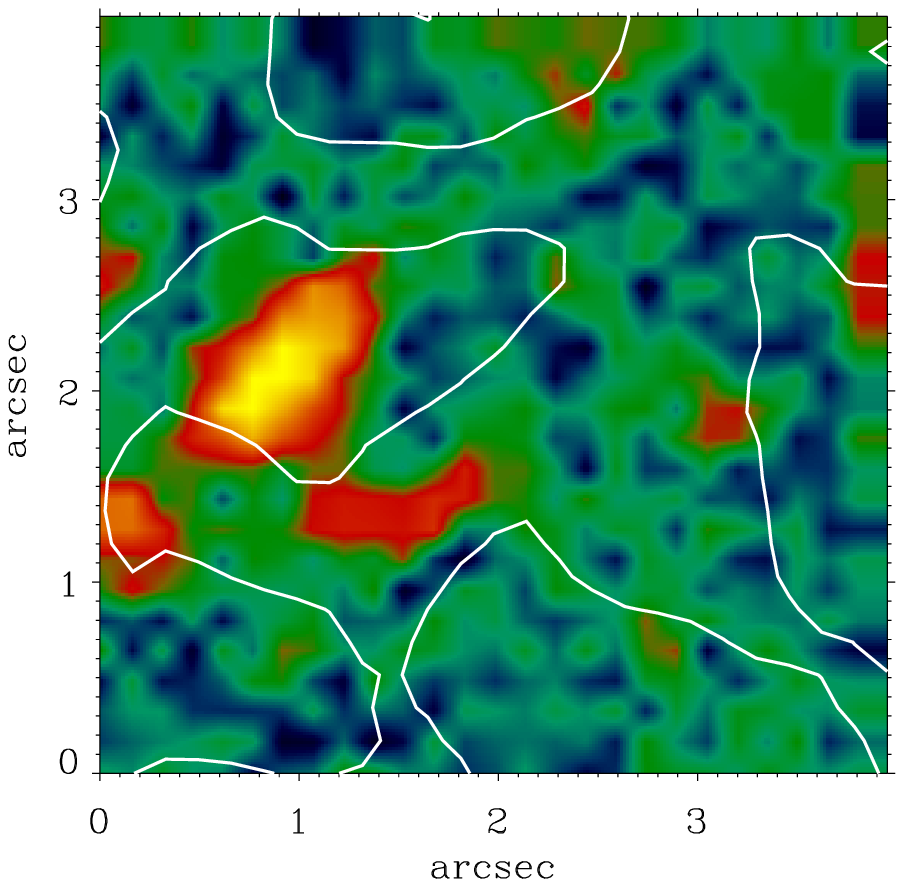}
\includegraphics[scale=.38]{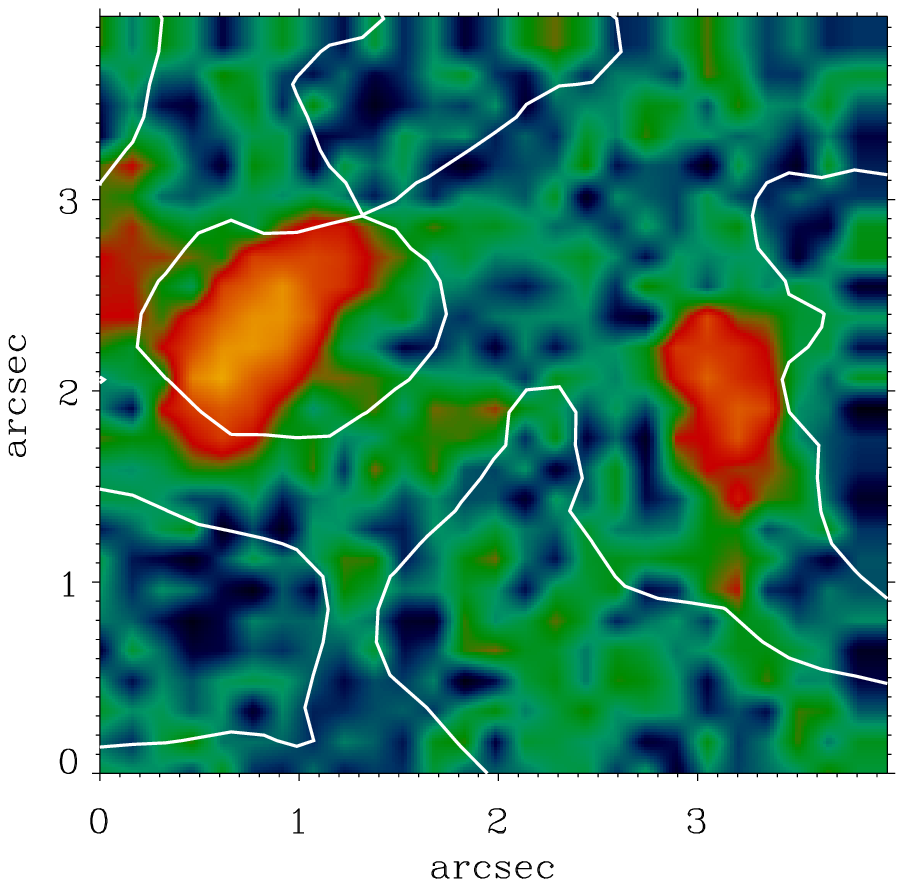}
\includegraphics[scale=.38]{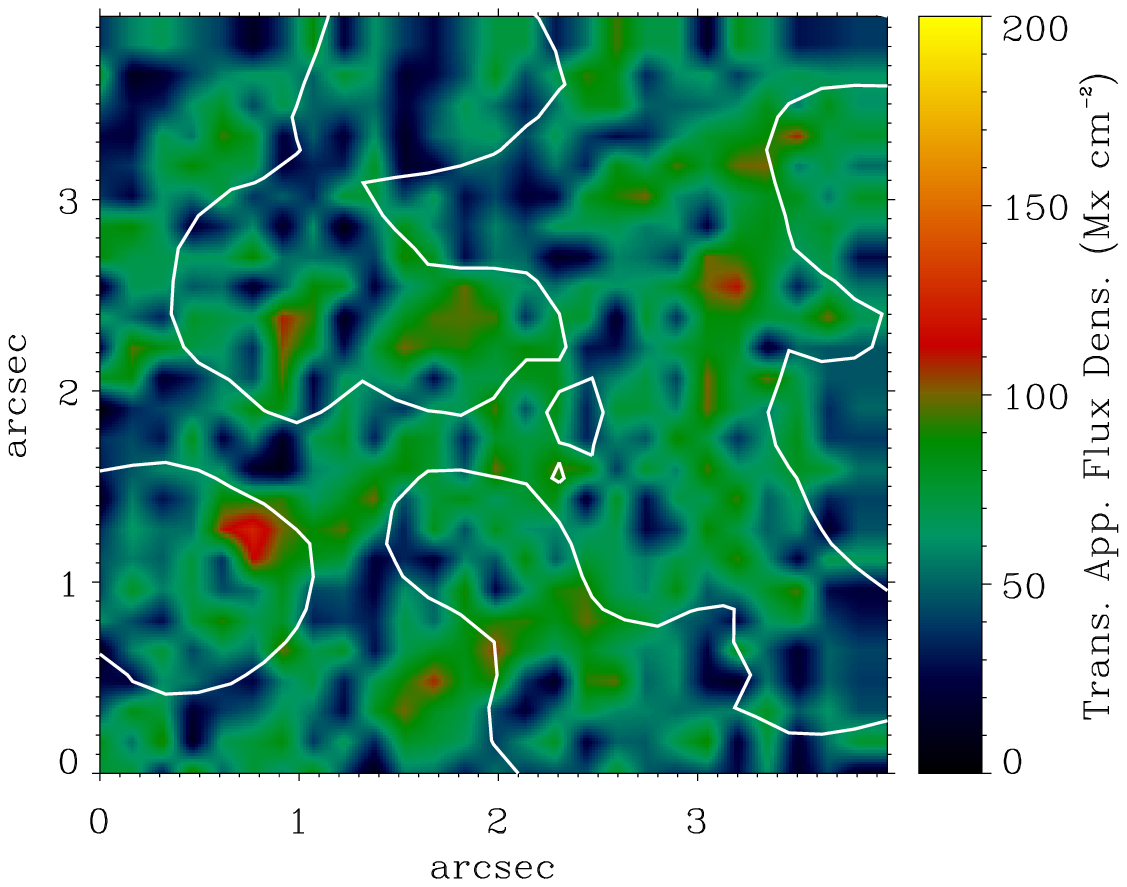}\\
\includegraphics[scale=.38]{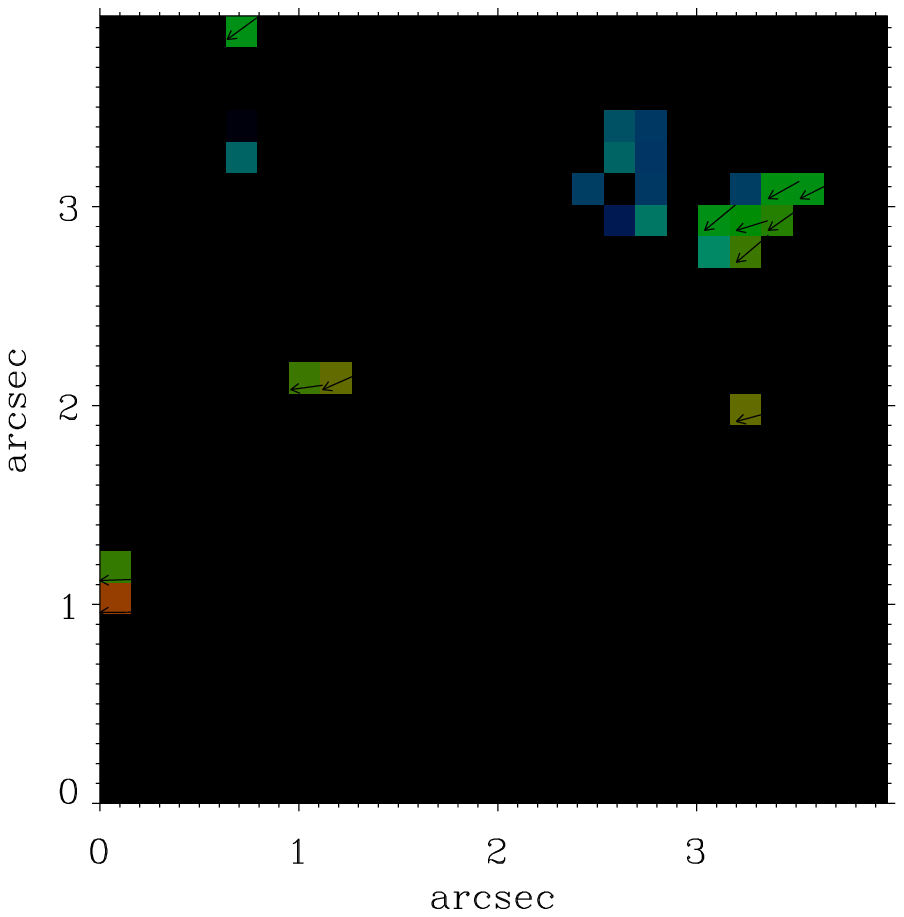}
\includegraphics[scale=.38]{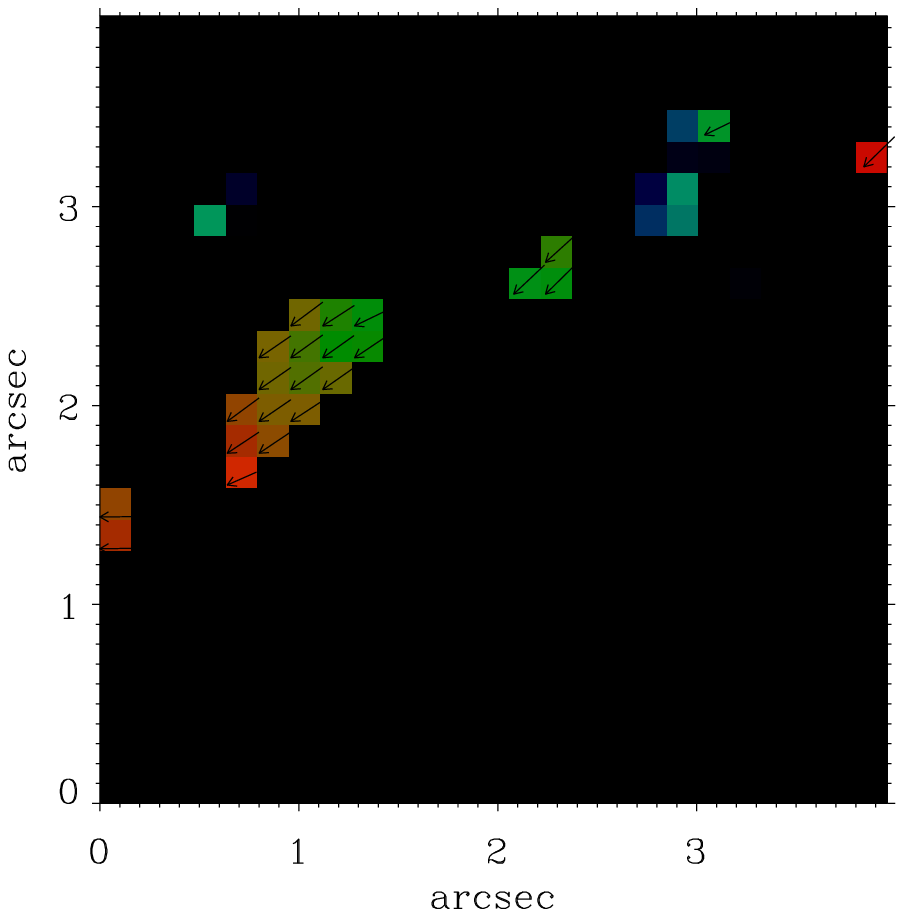}
\includegraphics[scale=.38]{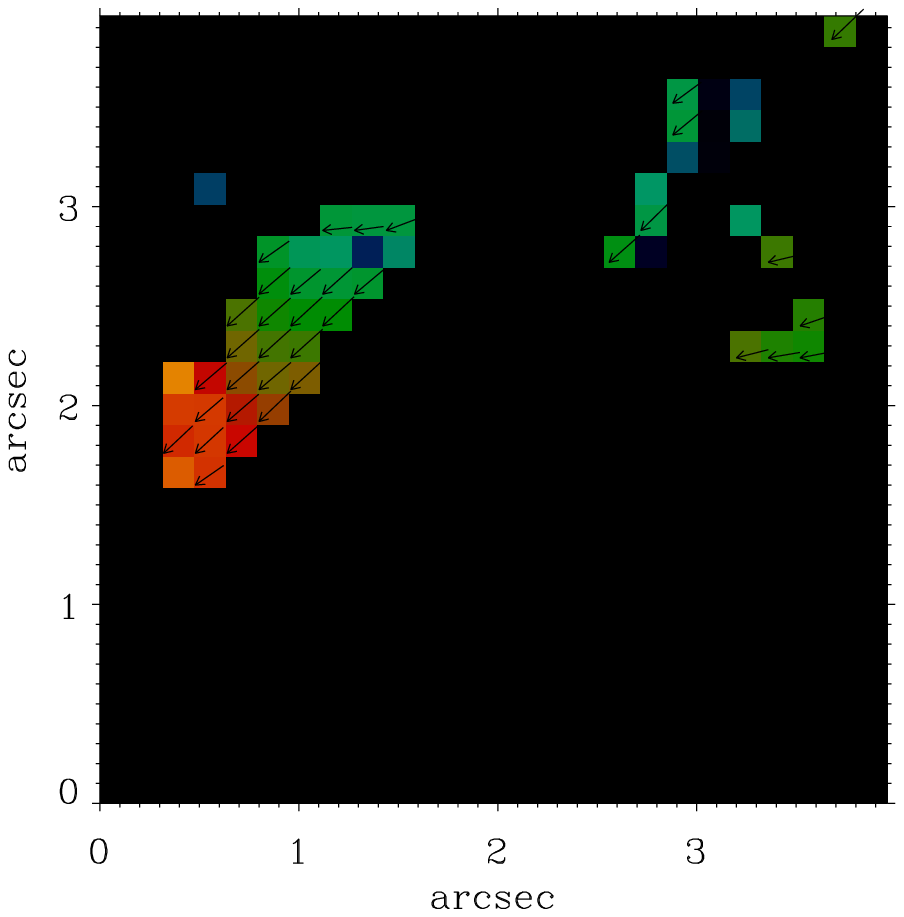}
\includegraphics[scale=.38]{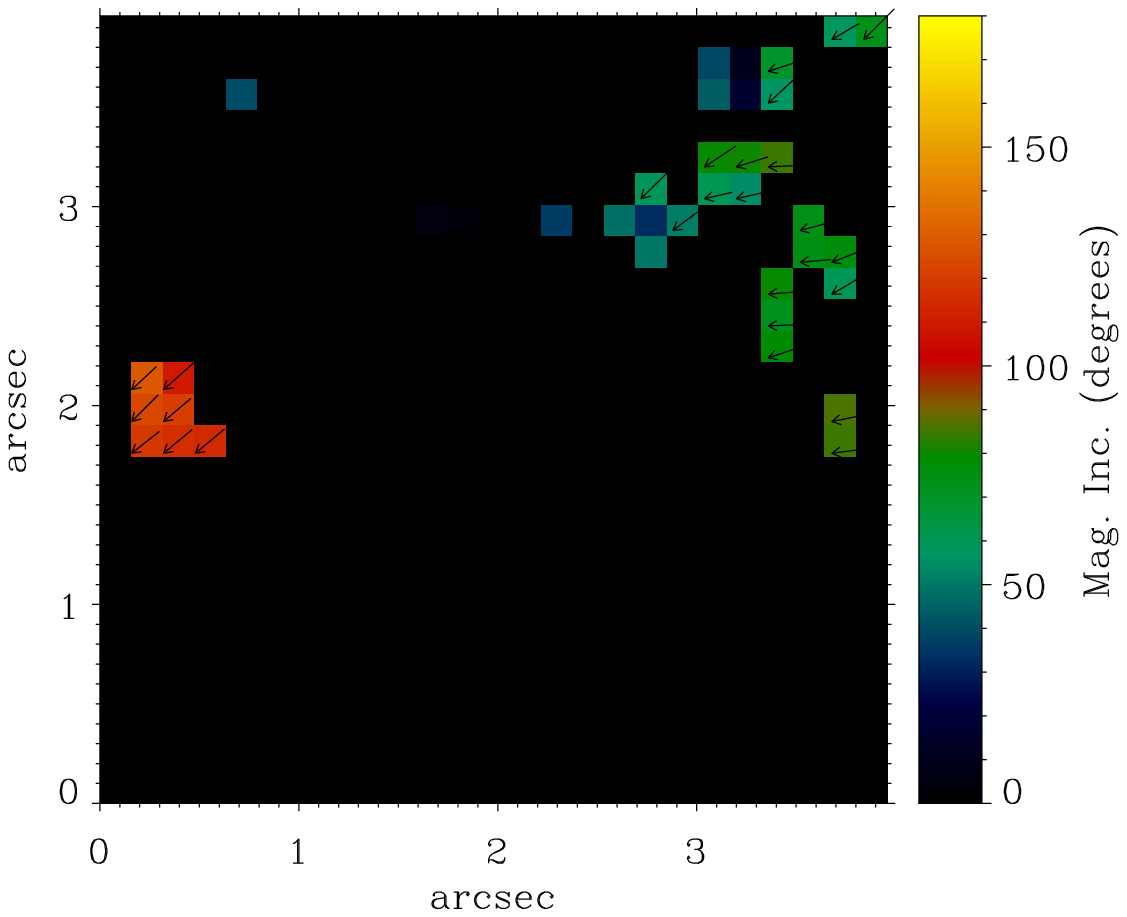}
\caption{Magnetic field properties of the region showed in 
Fig. 1. Time increases from left to right. The top and middle rows show the 
evolution of the apparent longitudinal and transverse magnetic flux densities 
respectively (white contours delimit the granular-intergranular boundaries). 
The bottom row shows the magnetic field inclination 
(color coded) and its azimuth (arrows) yielded by the LTE inversions.}
\end{figure}

In order to quantify the actual magnetic flux density and its topology, we 
carried out full Stokes LTE inversions (with LILIA, \cite{hector00}) of those 
pixels with non-negligible linear or circular polarization signals. 
While the temperature and the line of sight velocity were allowed to vary 
with 4 and 2 nodes respectively, the rest of the physical magnitudes 
were assumed constant during the inversion process. A fixed stray light 
profile (obtained from an average of non-magnetic data profiles) accounted for 
the non-magnetic atmospheric component.

\noindent Intrinsic field strength measurements in small-scale unresolved fields
using the Fe {\sc i} line pair can be rather uncertain for such weak
concentrations (\cite{marian06}), although \cite{orozco07} show that this 
seems to be a problem arising from poor spatial resolution. 
LTE inversions should give reliable values for the magnetic flux 
density. However, some of the polarization signals here are
marginally above the noise level, and different trial inversions performed 
with various assumptions (forcing the magnetic field to be constant with 
height or allowing for a linear variation, fixing the filling factor to 1 or
allowing for a free amount of stray light..) resulted in different values of 
the retrieved flux density, although the fitting performance of the code
is comparable in all cases.
For this reason we decided to compute the apparent longitudinal and 
transverse magnetic flux densities directly from the integrated polarization
signals (without further model assumptions) as explained in \cite{lites07a}, 
but we adopted the inversion results concerning the magnetic field topology
(which remains consistent independently of the assumptions). 

The top rows of Fig. 2 show the computed longitudinal and transverse 
apparent flux densities in the 4"x4" box throughout the time sequence. 
The white contours delimit the granule-intergranule boundaries.
The bottom row of Fig. 2 represents the field orientation as retrieved by 
the LTE inversions, with color-coded pixels accounting for the inclination 
values and arrows showing the direction of the magnetic field azimuth. 
Although the 180 degree ambiguity is not resolved, the arrows were drawn to 
point towards the positive polarity. 
For $t=0$ s (boxes in the first column) there is barely any magnetic signal in 
the center-left side of the box. Only 2 minutes later (second column), a new 
concentration of mostly horizontal magnetic flux appears. 
The magnetic field is parallel
to the solar surface and its azimuth makes an angle of $\sim 60$ degrees with 
the E-W direction. In the third snapshot ($t=250$ s), this magnetic "blob" has 
stretched in the linear direction and clearly developed into
a loop-like structure, with two opposite magnetic poles connected
by horizontal magnetic field. 
The total new longitudinal magnetic flux measured at the footpoints of this 
loop turns out to be $\sim 10^{17}$ Mx.
In the last snapshot (6 minutes after the first one), the central part of 
the loop is not detectable anymore, but the vertical dipoles still remain
visible, although they have drifted away towards the boundaries of the granule
from which they emerged.

Due to the azimuth ambiguity there are two possible topology configurations
(sketches A and B in Fig. 3) for the magnetic loop seen in the third column 
of Fig. 2.
If we consider the facts that the plasma motion is upward inside
the granule (the spectral profiles are clearly blueshifted) 
and that the horizontal magnetic flux is the first to appear 
-followed by the vertical dipoles- the most likely picture in this particular
case is the one presented in cartoon B of Fig. 3. 
We can understand the whole emergence event as follows: as the loop rises
in the atmosphere, its top part reaches the surface before its footpoints
can make it to the same level (see sequence B in Fig. 3).
 The expansion of the structure as it develops, 
together with the fact that it seems to be reaching higher layers in the 
atmosphere, makes the horizontal magnetic signal in the center of the loop 
undetectable to the Fe {\sc i} lines at the end of the sequence. 

\begin{figure}[t]
\includegraphics[scale=.6]{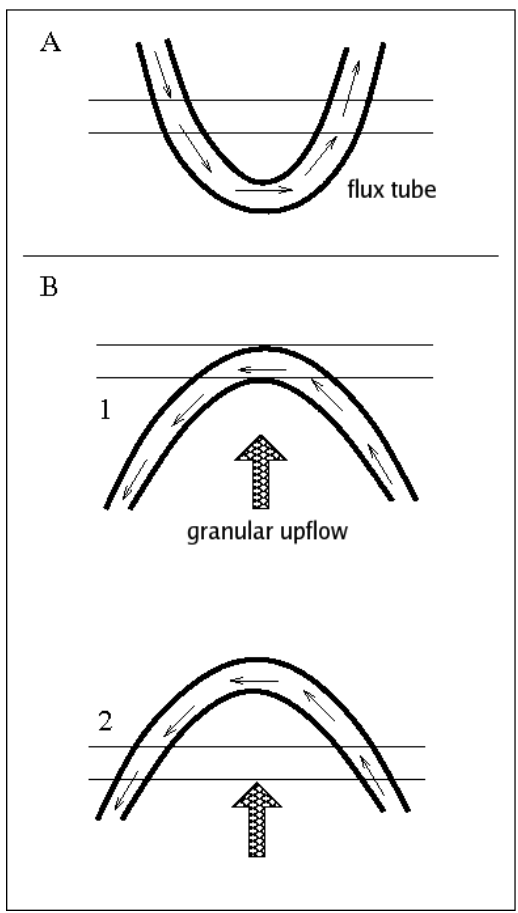}
\caption{Possible magnetic field configurations due to the azimuth ambiguity.
Horizontal parallel lines represent the solar surface. Sketch B represents
the evolution of the most likely configuration, as the flux tube arises 
above the surface pulled by the upward granular plasma motions.}
\end{figure}

\section{Conclusions}

We present observational evidence for an emerging small-scale magnetic loop
structure in the quiet Sun disk center. 
For the first time, the full topology of an emerging magnetic loop in 
the quiet Sun IN was followed in time from the moment it showed the first hints
of a signal until it was fully developed. Magnetic flux emerges within a 
granular region, showing strong horizontal magnetic signal flanked by
traces of two vertical opposite polarities on each side of it.  
As time goes on, the traces of the horizontal field disappear while the the 
vertical dipoles drift -carried away by the plasma motions- towards the 
surrounding intergranular lanes, where they stay trapped for a while 
and aggregate to other magnetic field concentrations resulting in larger 
flux elements.
This emergence event brings to the surface $\sim 10^{17}$ Mx of 
apparent longitudinal magnetic flux and does not seem to have any significant
influence on the shape of the underlying granulation pattern. This
is in agreement with the simulations presented in \cite{cheung07}, where
small-scale flux tubes with less than $10^{18}$ Mx of longitudinal flux
are not sufficiently buoyant to rise coherently against the granulation, 
and produce no visible disturbances in it.

From this example it seems that, although the flux emerges cospatial with 
a granule, the time that this small loop stays inside the granular structure 
is typically small (several minutes). The convective motions carry the
vertical magnetic flux towards the intergranular lanes, where it stays 
confined for longer times. Similar studies on magnetic flux from remanent
active regions (\cite{ishikawa07}) point to the same conclusions. 
This could explain why the transverse magnetic
flux observed at disk center is, in general, co-spatial with granules while 
the longitudinal flux tends to be concentrated in the intergranular lanes
(\cite{lites07b}). This idea is also in agreement with \cite{harvey07},
who showed that the seething horizontal fields they measured with the GONG 
network instruments and the vector spectromagnetograph of SOLIS (VSM) seem 
to be driven by granular and supergranular convection. According to 
these authors, the absence of 
longitudinal and latitudinal dependence on the full disk measurements favors 
the idea that these fields are created and destroyed by local processes.

\acknowledgments
We wish to thank all those involved in the international Solar-B/Hinode
program for their dedication and work over many years. We also acknowledge
those members of the Hinode/SOT team not listed specifically as authors
on this paper who helped in the planning of the observations presented
herein.
Hinode is a Japanese mission developed and launched by ISAS/JAXA, with
NAOJ as domestic partner, and NASA and STFC (UK) as international partners.
It is operated by these agencies in co-operation with ESA and NSC (Norway).
The FPP project at LMSAL and HAO is supported by NASA contract 
NNM07AA01C.

%% the file, uncomment the \plotone, \plottwo, and

%\begin{figure}
%\includegraphics[angle=90,scale=.50]{f3.eps}
%\caption{Animation still frame taken from \citet{kim03}.
%This figure is also available as an mpeg
%animation in the electronic edition of the
%{\it Astrophysical Journal}.}
%\end{figure}

\end{document}